\documentclass[global,twocolumn]{article}
\usepackage{graphics}
\usepackage{amsmath}
\usepackage{amssymb}
\usepackage{xcolor}
\usepackage{url}

\usepackage[square,numbers,sort&compress]{natbib}
\begin{document}
\title{Efficient and robust photo-ionization loading of beryllium ions}
\author{Sebastian Wolf$^1$, Dominik Studer$^1$, Klaus Wendt$^1$, Ferdinand Schmidt-Kaler$^1$\\ \\ $^1$QUANTUM, Institut f\"ur Physik, Johannes Gutenberg Universit\"at Mainz}              
%
%
\maketitle
\begin{abstract}
We demonstrate the efficient generation of Be$^+$ ions with a 60\,ns and 150\,nJ laser pulse near 235\,nm for two-step photo-ionization, proven by subsequent counting the number of ions loaded into a linear Paul trap. The bandwidth and power of the laser pulse are chosen in such a way that a first, resonant step fully saturates the entire velocity distribution of beryllium atoms effusing from a thermal oven. The second excitation step is driven by the same light field causing efficient non-resonant ionization. Our ion-loading scheme is more than 15 times more efficient as compared to former pathways using two-photon continuous wave laser excitation. 
\end{abstract}
\section{Methods of photo-ionization and modern applications}
\label{intro}

Robust and efficient ionization schemes are essential for many applications: Trapped atomic ions of the species Ca$^+$, Be$^+$, Sr$^+$, Yb$^+$, or Mg$^+$ are employed for quantum computing \cite{leibfried2003experimental, schmidt2003realization, moehring2007entanglement, morigi2001two}. In such experiments, ions and ion crystals are trapped and operated as qubits in micro structured sophisticated Paul trap devices \cite{kaufmann2017scalable}. Any contamination or charging of the nearby trap electrode surfaces affects the motional heating rate, which is relevant for the quality of the quantum processes. Similarly, trapped single ions are employed in fundamental precision experiments for frequency standards \cite{ludlow2015optical}, and again, slightest electric stray fields or their drifts result in micro motion effects which lead to systematic uncertainties of the clock transition \cite{keller2016evaluation, keller2015precise}. Actually, the largest uncertainty of the frequency standard of the best Yb$^+$ single ion clock standard results from the second-order Doppler shift caused by the residual secular and micromotion of the ion \cite{huntemann2016single}. Also if single ions are used for preparing ultra-cold antimatter atoms for tests of fundamental symmetries, any uncompensated or drifting stray charge would affect the accuracy of this measurement. Consequently, quite some attempts have been made in the past to improve the generation of ions and their loading into Paul traps and to prevent any uncontrolled contamination or charging of the trap structure as far as possible. While in the past, ion traps have been loaded by an inefficient electron-beam bombardment method for ionization of atoms from a neutral atomic beam or residual gas reservoir, the demonstration of optical resonant two-step photo-ionization was a great achievement \cite{kjaergaard2000isotope}. In such way, the charging of nearby trap electrodes was significantly reduced and furthermore the scheme became selective to load only one desired atomic species and isotope. Using UV laser diodes for this task \cite{gulde2001simple} and solid state laser systems \cite{lo2014all,blakestad2010transport} made the method applicable for many different atomic ions. However, with the miniaturization of segmented ion traps, the available trapping volume shrinks and so does the loading rate, if the flux of neutral atoms is not increased. In this scenario even the slightest surface contamination leads to an increased motional heating rate
of the trapped ions and also drifting electric potentials \cite{daniilidis2014surface,hite2012100}. Typically, small apertures are employed to reduce the amount of deposited material from the oven on trap electrode surfaces. Nevertheless, the contamination of electrode surfaces by the effusive neutral atom beam is hard to reduce, or even fully mitigate, especially in the case of ion trap devices with $ \leq$\,100\,$\mu$m dimensions.

 A second strong motivation for our work is found in the field of exotic ions, i.e. in view of quantum logic spectroscopy for the purpose of ion clocks \cite{windberger2015identification} or to test long-term variations of fundamental constants \cite{uzan2003fundamental, derevianko2012highly}. A specifically efficient photoionization method is required to load such ions. In this paper we apply a pulsed resonant laser ionization process for loading $^9$Be$^+$ ions, the particular species that has been widely employed for quantum logic operations, sympathetic cooling and quantum logic spectroscopy. The pulsed laser multi-photon ionization scheme had been previously optimized for selective laser-mass-spectrometric applications \cite{ledingham1997high}. It provides high and well controlled ionization probability even along weak optical resonances and prevents the drawbacks mentioned before. First, the laser source and experimental apparatus are outlined, then measurement results on loading $^9$Be ions into the microtrap are shown. Finally, we discuss the advantages of the new method and its applicability for loading other ion species of relevance for quantum information processing with ion-qubits or high precision spectroscopy. 


\section{Pulsed laser source near 235\,nm}
\label{sec:laser}

For resonant photoionization of beryllium atoms, we use the fourth harmonic of a high-repetition rate pulsed\\ Ti:sapphire laser. That particular laser type in use is a standard component of the majority of laser ion sources at on-line isotope production facilities worldwide \cite{yi2003temporal,rothe2011complementary}. As pump source serves a high repetition rate pulsed, frequency-doubled Nd:YAG laser\footnote{Clark-MXR Orc-1000} with pulse lengths of 300-500\,ns and a beam quality factor of $M^2 < 30$ at a repetition rate of 7\,kHz.

The Ti:sapphire laser is based on a symmetric Z-shaped standing-wave cavity with a highly focused central arm, formed by two concave mirrors $(r= 75\text{\,mm})$, which are placed at a folding angle of 17.75$^\circ$ around a central, Brewster-cut Ti:sapphire crystal\footnote{Roditi International Corporation} with length 20\,mm. The crystal is positioned at a distance of 20\,mm and 50\,mm to the concave mirrors. The two outer arms provide widely parallel beams for the installation of wavelength selective elements with foci on the high reflector and the output coupler $(R=80\text{\,\%})$, respectively. Correspondingly, their precise length is non-critical and a resonator round-trip  length in the range of 495\,mm is typically realized, resulting in a free spectral range of $\approx 300$\,MHz. This arrangement serves for full compensation of the astigmatism induced by the Brewster cut elements in the laser cavity. The pump beam is focused into the Ti:sapphire crystal through one of the concave mirrors by a 90 mm plane-convex pumping lens to a waist size of 100\,$\mu$m slightly exceeding the resonator mode waist of about 40\,$\mu$m. Frequency selection is achieved by a 3-plate birefringent filter and a 0.3\,mm etalon $(R=40\text{\,\%, }\text{FSR}=325\text{\,GHz})$ installed in one of the outer parallel arms of the cavity. With the resulting fundamental linewidth of 5 - 8\,GHz most hyperfine splittings, isotope shifts and Doppler classes in thermal velocity distributions are covered, allowing efficient ionization of an atomic species of interest. Optionally, one may insert an additional etalon  with a thickness of 5\,mm and $R=8.5\text{\,\%}$ to further narrow down the laser linewidth to 1 - 2\,GHz \cite{sonnenschein2014characterization}. The laser can be operated\footnote{with a total of four different laser mirror sets} from 680\,nm to 960\,nm with a mode-hop-free tuning range of about $325$\,GHz (one etalon FSR). Depending on pump power and wavelength the average output power of tunable laser radiation can reach up to 5\,W. The pulse lengths are typically between 40\,ns and 60\,ns. For the work discussed here, we operate the laser with 12\,W pump power generating typically 1.5\,W of output near 940\,nm.

From the fundamental laser radiation at 940\,nm we generate UV in two externally installed consecutive single-pass second-harmonic generation stages by strong focusing of the beam into BBO crystals of length 8\,mm with cutting angles of $\theta=25.2 ^\circ$ and $\theta=58.2 ^\circ$, respectively\footnote{Crysmit Photonics Co}. With a conversion efficiency of 5 to 10\,\% for each stage, an average output power of 1.5\,mW is available at 235\,nm. Note, that in comparison to cw radiation the pulsed nature of the radiation leads to a photon flux during the laser pulse which is governed by the duty cycle of $\approx50\,\text{ns}/140\,\mu$s, corresponding to an enhancement of a factor of 3000. The spectral linewidth is enlarged by a factor of about $\sqrt{2}$ in each frequency doubling process. In addition it should be mentioned that for a pulsed laser source the extension of the wavelength range with frequency doubling, or third harmonic generation, is straightforward, does not require any kind of enhancement cavities or even careful alignment and works efficiently and reliably. In the case of our setup, a spectral range from 215\,nm to 480\,nm is accessible by non-linear conversion processes of the fundamental laser output. 

\section{Segmented micro-structured ion trap and ion detection system}
\label{sec:trap}

A segmented linear Paul trap is used for the experiments, see Ref.~\cite{jacob2016transmission} for details. It consists of four gold-coated and micro-structured alumina wafers and two titanium endcaps with 600\,$\mu$m holes through which the laser radiation for in-trap ionization is sent. The ion-electrode distance is 470\,$\mu$m and the ion-endcap distance is 1.45\,mm. The RF drive-frequency is $2\pi \times 30$\,MHz with $U_\text{peak}$ = 50\,V, resulting in trap frequencies\\ $\omega_{r1, r2, z}/(2\pi) = (2.060, 3.070, 1.400)$\,MHz for the $^9$Be$^+$-ions. The atoms are provided by a 5\,mm long heated beryllium wire with a diameter of 250\,$\mu$m. It is glued into one of four holes in a 5\,mm long round ceramic tube. Through the remaining three holes a tantalum wire is wound and electro thermally heated by a current of 2.3 - 2.4\,A. This oven is located 20\,mm away from the trap center and has an angle of $90^\circ\pm 5^\circ$ to the trap axis and the ionization laser beam.

\begin{figure}
\resizebox{0.45\textwidth}{!}{%
  \includegraphics{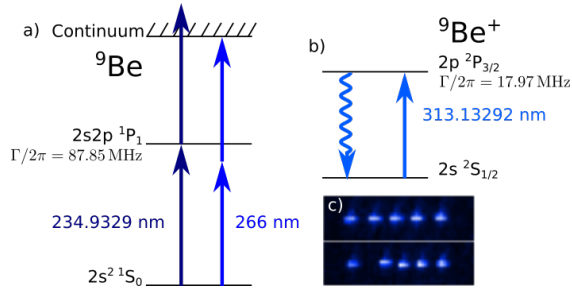}}
\caption{Levels and transitions used for a) resonant excitation at 235 nm and subsequent non-resonant ionization in neutral beryllium, for non-resonant two-photon ionization at 266 nm and for b) Doppler cooling and fluorescence detection in the beryllium ion, hyperfine and Zeeman sublevels not shown. c) A single-species ion crystal and a crystal with a dark impurity, exposure time 50\,ms.}
\label{fig:1}      
\end{figure}

Beryllium atoms emerging from the oven are subsequently ionized, trapped, cooled and the emitted fluorescence is detected on a CCD camera  with single ion sensitivity. For Doppler cooling we employ the $2s\ ^2 S_{1/2} \rightarrow 2p\ ^2 P_{3/2}$ transition near 313\,nm, see Fig. \ref{fig:1}b. This wavelength is generated using a DBR laser-diode at 626\,nm \cite{blume2013monolithic, cozijn2013laser} as master oscillator with 13\,mW power, injected into a 170\,mW slave diode \footnote{Thorlabs, HL63133DG}. Injecting a power of 70\,mW into a power enhancement cavity with parameters $\mathcal{F}=260$ and $\text{FSR}=1.5$\,GHz for frequency-doubling results in a power of up to 500\,$\mu$W at 313\,nm wavelength. The Master laser frequency is permanently locked to a wavelength meter\footnote{Toptica, HF-ANGSTROM WS/U-30U} ensuring a long-term frequency stability of 1.25\,MHz. For phase modulation of the beam near 313\,nm we are using an electro-optical modulator (EOM) and generate sidebands at 1.25\,GHz with a modulation index of about 0.9\,rad. One sideband is resonant to the $2s\ ^2 S_{1/2}\,(F=1) \rightarrow 2p\ ^2 P_{3/2}$ transition, required for repumping population from the F=1 hyperfine groundstate level to achieve continuous fluorescence. The laser beam is transmitted through a quarter wave plate for circular polarization and focused onto the ions with a $f=200$\,mm lens, with a k-vector parallel to the magnetic field direction and a non-vanishing projection on all three trap axes. 

For detecting the ions a $f/1.6$ objective images the fluorescence light near 313\,nm onto an EMCCD camera with a magnification of about 15. This arrangement results in a clear distinction in neighboring ions as long as the ion number is smaller than about 6, restricting the number of ions for each measurement cycle, as shown in Fig. \ref{fig:1}c.

\section{Experimental demonstration of efficient photoionization}
\label{sec:results}
The experiment involves a three-step sequence: a series of 700 ionization laser pulses is fired. If the $2s^2\ ^1S_0\rightarrow 2s2p\ ^{1}P_1$ transition near 235\,nm is excited a photon from the same pulse may ionize the atom,  as shown inFig. \ref{fig:1}a. Ions in the trapping volume are captured and, as the light field near 313\,nm is continuously turned on, they are cooled and almost instantaneously fluorescence is observed on the CCD camera. We have chosen a cooling time of 20\,s, then the number of ionized, loaded and cooled ions is counted. Finally, the trap is emptied by switching the RF drive off. This measurement cycle is repeated 120 times, so that each data point consists of a total of 84000 laser pulses. 

\begin{figure}
\resizebox{0.45\textwidth}{!}{%
  \includegraphics{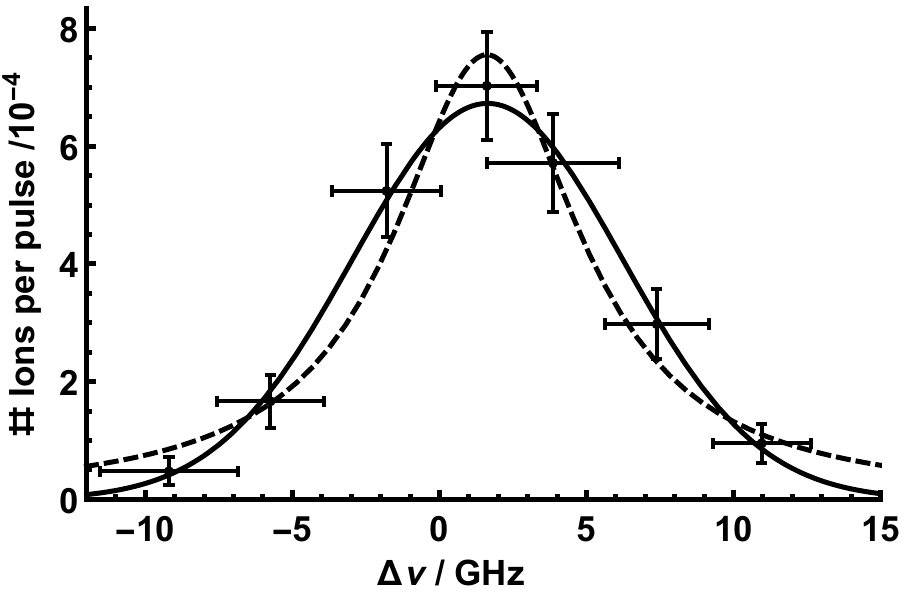}}
\caption{Ionization efficiency versus laser frequency detuning from the NIST value. The errors are given by Poissonian counting statistics and standard deviations of the wavemeter logs. The data is fitted with a Lorentzian distribution (dashed) and a Gaussian distibution (solid). The Gaussian fit gives a FWHM linewidth of  $2\pi\times 10.8(7)$\,GHz and a detuning of $2\pi\times 1.6(4)$\,GHz. The line shift can be fully explained by the Doppler shift of about 0.9\,GHz. The peak pulse power for this measurement was 0.72\,W.}
\label{fig:2}       
\end{figure}

For obtaining information about the resonance behaviour of the photoionization process, we scan the Ti:Sa-laser frequency with the intra-cavity etalon while we determine its wavelength with the wavelength meter. The dependency is given in Fig. \ref{fig:2}. A systematic offset of $2\pi\times 1.2(4)$\,GHz of the wavemeter readout was determined from independently performed calibration measurements and is subtracted in the data. The frequency errors are the standard deviation in the logged wavemeter data. Data points near the maximum account for about 60 ionization events, and the errors in the number of ionized atoms are derived from Poissonian statistics. 

The Gaussian fit to the data in Fig.~\ref{fig:2} reveals a FWHM linewidth of $2\pi\times 10.8(7)$\,GHz and a detuning of $2\pi\times 1.6(4)$\,GHz. The natural linewidth for this transition is $2\pi\times 88$\,MHz. The melting point of beryllium is 1560\,K, assuming that temperature and a $5^\circ$ angle between the atom beam and the laser beam, a FWHM Doppler broading of only $2\pi\times 1.3$\,GHz is estimated. This means the measured linewidth is fully dominated by the linewidth of the frequency quadrupled laser radiation with a FWHM of up to $2\pi\times 14$\,GHz \cite{rothe2011complementary}. The Doppler shift for the given temperature is $2\pi\times 900$\,MHz which is consistent with the measured line shift. 

\begin{figure}
\resizebox{0.49\textwidth}{!}{
  \includegraphics{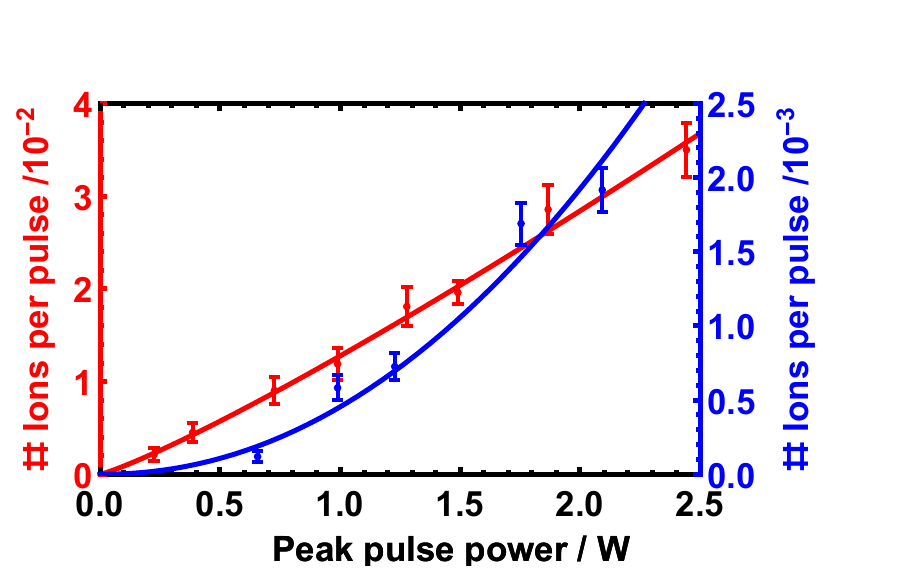}}
\caption{Ionization efficiency versus peak pulse power for the resonant pulsed ionization scheme. Errors are calculated from Poissonian counting statistics and data are given for ionization on resonance (red points) and with a 3.3 FWHM linewidth red detuned laser (blue points). Note the different scalings on the y-axis for both cases. Power-law fits reveal an exponent of 1.15(8) and 2.08(15), respectively. No background ionization event were observed without laser irradiation. 
}
\label{fig:3} 
\end{figure}

To validate the nature of the  ionization as a two-photon process, the ion loading rate is measured as function of the peak pulse power. For excitation at the maximum of the resonance in Fig.~\ref{fig:2}, thus fully resonant with respect to the first excitation step, we find a linear slope, as indicated in Fig. \ref{fig:3} as red data points and red fit curve. The power-law fit reveals an exponent of 1.15(8) consistent with a linear dependency, somehow surprising for a two-photon process at first glance. We additionaly test the scheme more carefully, with the excitation laser frequency  detuned to the red side of the resonance by $\Delta= 3.3\,\Gamma_{\text{FWHM}}$. Now, we observe an orders of magnitude lower excitation rate. In order to obtain a good signal with sufficiently many detected ions, here the oven temperature was increased for this measurement to 2.4\,A. Results are given in Fig. \ref{fig:3} as blue data points and blue fit curve. Notably, now the power-law fit exponent results in 2.08(15) confirming an expected quadratic behavior. Therefore, we conclude that in the case of resonant excitation by the pulsed laser (Fig. \ref{fig:3} red data points) the first step from the beryllium atomic ground state 2s$^2$\, $^1$S$_0$ to the 2s2p $^1$P$_1$ state is {\em fully saturated} in the range of laser pulse powers of our experiment and the observed linear increase in loaded ions arises entirely from the power dependency of the second non-resonant excitation step into the continuum.

\begin{figure}
\resizebox{0.55\textwidth}{!}{%
  \includegraphics{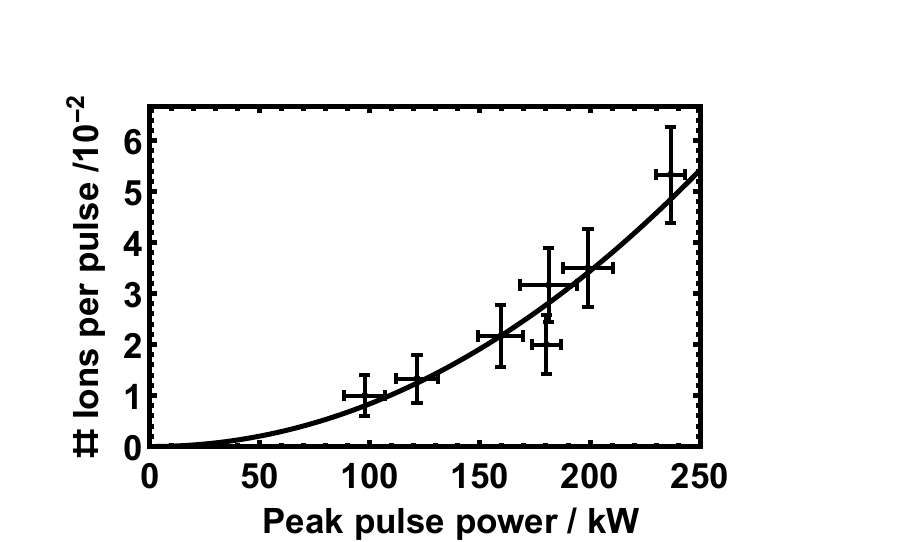}}
\caption{Ionization efficiency for the non-resonant two photon process. The errors are given by Poissonian counting statistics and the standard deviation of the pulse power. The data is fitted with a power-law reveiling an exponent of 2.0(5).}
\label{fig:4}       
\end{figure}

We compare this highly efficient pulsed resonant scheme with a pulsed off-resonant ionization scheme, where we use the fourth harmonic of a commercial Nd:YAG pulsed laser near 266\,nm, which is also indicated in Fig. \ref{fig:1}a. The energy of two photons at this wavelength coincides with the ionization energy of beryllium up to a per mille. Now pulses have a duration of about 5\,ns and a pulse energy of up to 1.2\,mJ compared to about 0.2\,$\mu$J for the resonant two step process. For measurements of the non-resonant ionization scheme we apply five laser pulses and count the number of trapped ions after a waiting and Doppler cooling time of 20\,s. To obtain good statistics, we repeat this 120 times for each data point, which corresponds to about a total of 40 times the photon number of the resonant two step ionization. The data for the dependency of the ionization rate against the peak pulse power is plotted in see Fig. \ref{fig:4}. The error bars for the power are the standard deviation in the power measurement over a minute. The error bars in the ionization probability are given by Possionian counting statistics. The data is fitted with a power-law  leading to an exponent of 2.0(5), which is consistent with the expected quadratic dependency for a non-resonant two-photon process. 

The foci of the laser at the trap center are measured with a CMOS camera moved through the attenuated focus\footnote{the beam diameters are calculated according to ISO~11145}. For the resonant laser the beam waist is 2400\,$\mu\text{m}^2$ and for the non-resonant laser it is  2470\,$\mu\text{m}^2$, both with about 10\,\% precision. Let us note, that the application of loading laser pulses did no lead to drifts in the micro-motion compensation voltages, even after hours of operation. This holds for both, the resonant laser near 235\,nm with a average power of 1.2\,mW and of the non-resonant laser near  266\,nm with average power of 12\,mW. The power values were either measured directly for the resonant laser or have been calculated from  the measured pulse energy and duty cycle for the non-resonant laser system.  


\section{Discussion of different photo-ionisation methods}
\label{sec:discussion}
We have applied an ionization method of resonant pulsed multi-photon laser excitation for loading of a linear Paul trap. Such a scheme is commonly used as Resonance Ionization Laser Ion Source (RILIS) at on-line isotope production facilities, because of its efficiency, its high mass and isotope selectivity, and because it works fairly general for a wide range of elements. Ref.~\cite{rothe2011complementary} lists 38 elements, for which RILIS was already tested and 23 elements for which efficient excitation and ionization schemes have been worked out, according to the data table \cite{smithatomic, rilis}. The pulsed Ti:Sa laser and the frequency conversion units are easy to adapt for the entire accessible wavelength range in the fundamental operation of 680\,nm to 1020\,nm, as well as between about 200\,nm to 500\,nm by non-linear frequency conversion. Additionally, the spectral bandwidth of the laser may be matched to the Doppler width of the atoms emerging from the oven to saturate the {\em entire velocity distribution} on the first transition, here for the case of the beryllium transition S$_0$ to P$_1$ near 235\,nm. In contrary to that, in previously used common cw resonant ionization scheme \cite{lo2014all}, the first excitation step will at most saturate atoms within the natural linewidth of $\Gamma/2\pi=88$\,MHz. In our case we determined a value for the Doppler width of about 1.3\,GHz, even though the oven beam is almost aligned at right angles with the excitation beam, because of the poor oven beam collimation. In most situations the Doppler width and shift will be even higher. From the comparison between natural and Doppler linewidth we estimate a $~$15-fold increase in efficiency of excitation on the first step. For the second excitation step to ionization the scenario of using a pulsed laser again improves the yield, as  efficient generation of the required UV radiation is easily achieved for the intense $\approx 50$\,ns laser pulses and reaches a significantly higher peak power density as compared to cw frequency doubling or UV laser diodes. Simultaneously, spatial and temporal overlap between the first and second step are optimal by design. In total we estimate a gain in loading efficiency by almost two orders of magnitude as compared to cw laser loading, while average laser light power for the high repetition rate pulsed laser system of about 2\,mW are similar to the ones used for resonant cw loading. Unfortunately, we can not provide a fully quantitative comparison in loading yield as we do not have an alternative cw photoionization system for beryllium in our lab \cite{lo2014all} besides the high power pulsed laser RILIS loading. 
The significantly higher loading efficiency allows for reducing the oven beam flux, thus mitigating contamination of trap electrode or insulator surfaces. As a results, we have not observed any drift in micromotion compensation voltages over all experiments with pulsed loading, which indicates that neither surfaces have  been charged up by photoelectrons nor contact potentials from contamination have been built up. Even when operating the linear segmented trap with gold-coated alumina wavers, we have not observed any damage. Correspondingly, for macroscopic traps, where surfaces are far from the loading region, even the non-resonant pulsed scheme applying high laser power density may be useful for beryllium, and could form a very cost effective alternative to the cw scenarios as used commonly to date.       


We acknowledge support from the DFG through the DIP program (Grant No. SCHM 1049/7-1), within the cluster of excellence PRISMA and from the EU through ENSAR2 RESIST (Grant No. 654002).

\bibliographystyle{apsrev4-1}
\bibliography{lit}%

\end{document}